\documentclass[aps, prd, amsmath, floats, floatfix, twocolumn, nofootinbib, showpacs]{revtex4-1}
\usepackage{graphicx}
\usepackage{color}

\newcommand{\bi}{\bibitem}
\newcommand{\be}{\begin{eqnarray}}
\newcommand{\ee}{\end{eqnarray}}
\newcommand{\rar}{\rightarrow}

\begin{document}

\title{Destroying the event horizon of regular black holes}

\author{Zilong Li}

\author{Cosimo Bambi}
\email[Corresponding author: ]{bambi@fudan.edu.cn}

\affiliation{Center for Field Theory and Particle Physics \& Department of Physics, Fudan University, 200433 Shanghai, China}

\date{\today}

\begin{abstract}
Recently, several authors have studied the possibility of overspinning or 
overcharging an existing black hole to destroy its event horizon and make 
the central singularity naked. When all the effects are properly taken into 
account, any attempt to destroy the black hole seems to be doomed to fail, 
in agreement with the weak cosmic censorship conjecture. In this letter, 
we study the possibility of destroying the event horizon of regular black holes. 
These objects have no central singularity and therefore they are not protected 
by the cosmic censorship hypothesis. Our results strongly support the 
conclusion that regular black holes can be destroyed. If we believe that 
the central singularity in astrophysical black holes is solved by quantum
gravity effects, we might have a chance to see the black hole's internal 
region and observe quantum gravity phenomena. As our finding implies the
violation of the black hole's area theorem, the collision of two black holes 
may release an amount of energy exceeding the Hawking bound, which
can be experimentally tested by gravitational wave detectors. 
\end{abstract}

\pacs{04.70.Bw, 04.20.Dw}

\maketitle

%%%%%%%%%%%%%%%%%%%%%%%%%%%%%%%

Under the main assumptions of the validity of the strong energy condition and 
of the existence of global hyperbolicity, in general relativity the final product of
the collapse is a singularity of the spacetime~\cite{sing}. At the singularity, 
predictability is lost and standard physics breaks down. According to the weak 
cosmic censorship conjecture, singularities produced in the gravitational 
collapse must be hidden within black holes and cannot be seen by distant 
observers~\cite{wccc}. The validity of this conjecture is still an outstanding 
open problem in gravitational physics. While it is thought to be correct at
some level, we also know physically relevant counterexamples in which naked 
singularities can be created from regular initial data~\cite{jm}. The validity of
the weak cosmic censorship conjecture is a fundamental assumption in the
theory of black hole thermodynamics. On the other hand, its possible violation
would allow to observe high curvature regions, where new physics is expected
to show up.

Some authors have studied the possibility of violating the weak cosmic 
censorship conjecture by destroying an existing black hole~\cite{w,h,js,gz,ung}. 
The advantage of this set-up is that we do not need any assumption 
about the matter content, which is the crucial and questionable point in
all the studies on the validity of the weak cosmic censorship hypothesis.
In 4-dimensional general relativity, a black hole is described by the Kerr-Newman
solution and it is specified by its mass $M$, spin $J$, and electric charge 
$Q$. The condition for the existence of the event horizon is
\be\label{eq-b}
M^2 \ge Q^2 + a^2 \, ,
\ee
where $a = J/M$ is the spin parameter. When Eq.~\eqref{eq-b} is violated,
there is no horizon and the central singularity is naked. One can thus imagine 
an experiment in which a black hole absorbs a small particle of energy $E$,
angular momentum $L$, and electric charge $q$. If the new object violates 
the bound~\eqref{eq-b}, the small particle has destroyed the black hole: the 
event horizon disappears and the central singularity becomes naked.

To see if it is really possible to destroy a black hole, we need to check if
the small particle with $E$, $L$, and $q$ can actually be absorbed by the 
black hole. In the test particle approximation, one can follow Ref.~\cite{w}
and finds that the {\it absorption condition} is
\be\label{eq-e}
E \ge \frac{q \left(A_\phi g_{t\phi} - A_t g_{\phi\phi}\right) 
- g_{t\phi} L}{g_{\phi\phi}} \, ,
\ee
where $A_\mu$ is the 4-potential of the electromagnetic field and 
Eq.~\eqref{eq-e} must hold from the point where the particle is fired to the
black hole's event horizon, where the particle is absorbed. If Eq.~\eqref{eq-e} 
does not hold, it simply means that the electric charge and/or the angular 
momentum of the test particle are too high and the particle cannot reach 
the event horizon. The {\it overcharging/overspinning condition} is instead
\be\label{eq-b2}
\left(M + E\right)^2 < \left(Q + q\right)^2 
+ \left(\frac{a M + L}{M + E}\right)^2 \, .
\ee

We can thus consider a near extremal black hole and find for which
values of $E$, $L$, and $q$ both Eqs.~\eqref{eq-e} and \eqref{eq-b2} are 
satisfied; that is, the particle can be absorbed and the new state has no
event horizon. In the Kerr-Newman spacetime in Boyer-Lindquist coordinates, 
the non-vanishing metric coefficients are
\be\label{eq-metric}
&&g_{tt} = - \left(1 - \frac{2 m r}{\Sigma}\right) \, , \quad
g_{t\phi} = - \frac{2 a m r \sin^2 \theta}{\Sigma} \, , \nonumber\\
&&g_{\phi\phi} = \left(r^2 + a^2 + \frac{2 a^2 m r \sin^2\theta}{\Sigma}\right) 
\sin^2\theta \, , \nonumber\\
&&g_{rr} = \frac{\Sigma}{\Delta}\, , 
\quad g_{\theta\theta} = \Sigma \, ,
\ee
while the 4-vector of the electromagnetic field is
\be
A_t = - \frac{Q r}{\Sigma} \, , \quad 
A_\phi = \frac{Q r}{\Sigma} a \sin^2 \theta \, ,
\ee
where
\be
\Sigma = r^2 + a^2 \cos^2\theta \, , \quad
\Delta = r^2 - 2 m r + a^2 \, ,
\ee
and $m$ is given by
\be
m = m_{\rm KN} = M - \frac{Q^2}{2 r} \, .
\ee
In Ref.~\cite{h,js,gz}, the authors find that a small particle can destroy a
black hole. However, these studies ignore the radiated energy as well as the
particle's self-energy. The former makes easier the destruction of the black hole, 
while the latter makes more difficult the absorption of the particle, as it may 
introduce a turning point before the particle reaches the black hole's event horizon. 
While there is not yet the theory and the technology to perform complete calculations,
analyses beyond the test-particle approximation strongly suggest that this is 
indeed the case and the self-energy acts as a cosmic censor~\cite{no-h,no-js}.

The validity of the weak cosmic censorship conjecture seems thus to be 
confirmed by these gedanken experiments when all the effects are properly 
taken into account, and it is apparently impossible to destroy an existing black 
hole. However, we may guess that a real black hole has no central singularity,
as the latter is more likely a pathological feature associated with classical
general relativity, to be removed by (unknown) quantum gravity effects. If this
is indeed the case, astrophysical black holes may not be protected by the
weak cosmic censorship conjecture and there may be a chance to destroy
their event horizon~\cite{hn}.

While we do not have yet any robust and reliable quantum theory of gravity 
capable of telling us how the singularities in the interior of black holes are solved, 
in the literature there are some toy-models of black hole solutions without the 
central singularity. The prototype of these regular black holes is the Bardeen 
metric~\cite{bar}, which can be formally obtained by coupling Einstein's gravity 
to a non-linear electrodynamics field~\cite{ag}. Another popular example is the 
Hayward black hole metric~\cite{hay}. The rotating solutions of the Bardeen 
and the Hayward metrics have been obtained in Ref.~\cite{rrbh}. In the non-rotating 
case, these solutions violate the strong energy condition, but not the weak one.
For spinning black holes, even the weak energy condition is violated, but such a 
violation can be made very small~\cite{rrbh}. In the rotating Bardeen and Hayward
black hole spacetime, the metric coefficients $g_{\mu\nu}$ are still given by 
Eq.~\eqref{eq-metric}, but $m$ is now given, respectively, by
\be
m_{\rm B} = \frac{M r^3}{\left( r^2 + g^2 \right)^{3/2}} \, , \quad
m_{\rm H} = \frac{M r^3}{r^3 + g^3} \, ,
\ee
where $g$ can be interpreted as the magnetic charge of the non-linear 
electromagnetic field or just as a quantity introducing a deviation from the Kerr 
metric and solving the central singularity. For neutral particles, the absorption
condition is still given by Eq.~\eqref{eq-e}. There is instead no simple
formula like Eq.~\eqref{eq-b} for the overcharging/overspinning condition,
so we have to check that the initial state is a black hole, i.e. the equation
$\Delta = 0$ (with the correct $m$) has at least one positive solution for $r$,
and that for the final state $\Delta = 0$ has no solutions.

\begin{figure*}
\begin{center}
\includegraphics[type=pdf,ext=.pdf,read=.pdf,width=17cm]{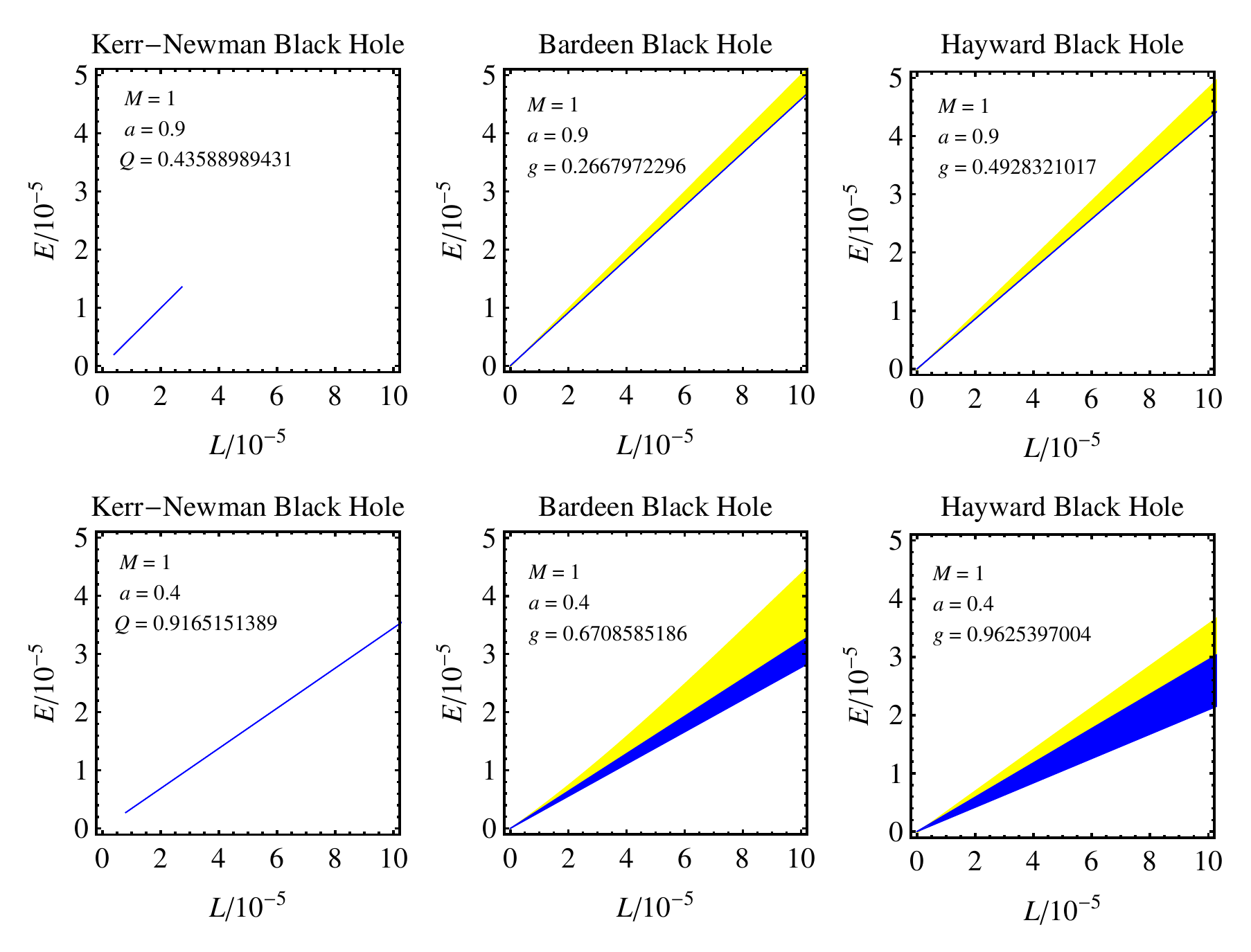}
\end{center}
\vspace{-0.7cm}
\caption{Values of the energy $E$ and of the angular momentum $L$ for which
a test particle can destroy the black hole. Left panels: Kerr-Newman black holes. 
Central panels: Bardeen black holes. Right panels: Hayward black holes. For 
regular black holes, the blue-dark region is for bound orbits, the yellow-light 
one is for unbound orbits. See the text for details.}
\label{fig1}
\end{figure*}

Let us now compare the attempt to destroy a Kerr-Newman black hole with the one 
for regular black holes. Since the nature of the electromagnetic fields is different, a
direct comparison is possible only considering the case in which the small 
particle has no charge. When we consider a near extremal Kerr-Newman
black hole\footnote{To be more specific, we first fix $a$ and then we take 
$Q = Q_{\rm extremal} - \varepsilon$ with $\varepsilon = 10^{-10}$. In the next
paragraph we do the same for the regular black holes, replacing $Q$ with 
$g$.}, we find that the region of the values of $E$ and $L$ for which the
small particle destroys the event horizon is very narrow. Two specific cases
are reported in the left panels of Fig.~\ref{fig1}. As noted in Ref.~\cite{gz,no-js},
the allowed energy range of $E$ is of order $L^2/M^3$. In our cases, we find
$\Delta E \sim 10^{-11}$, in perfect agreement with this estimate (here we 
have set $M=1$ and find $L \sim 10^{-5}$). The energy range of $E$ is of the 
same order (and opposite sign) of the expected corrections from the particle's 
self-force~\cite{gz,no-js}. It is thus plausible that the region disappears when 
this effect is properly taken into account and that the particle's self-energy 
prevents the destruction of the black hole, confirming the validity of the weak 
cosmic censorship conjecture.

We can then repeat the same experiment for the Bardeen and Hayward black 
hole. Specific cases of the region for which the test particle can destroy the 
event horizon are reported in the central (Bardeen) and right (Hayward) panels 
of Fig.~\ref{fig1}. The result is clear: now such a region is definitively larger than
the one in the Kerr-Newman case. In the plots in Fig.~\ref{fig1} we have also 
distinguished the bound (blue-dark region) and the unbound (yellow-light region) orbits 
capable of destroying the black hole. Since the particle's self-force should here 
introduce a correction of the same order of the Kerr-Newman case, $\sim L^2/M^3 
\sim 10^{-10}$, while the allowed region is significantly wider, $\Delta E \sim 10^{-5}$, 
we may conclude that regular black holes can be destroyed.

\begin{figure*}
\begin{center}
\includegraphics[type=pdf,ext=.pdf,read=.pdf,width=17cm]{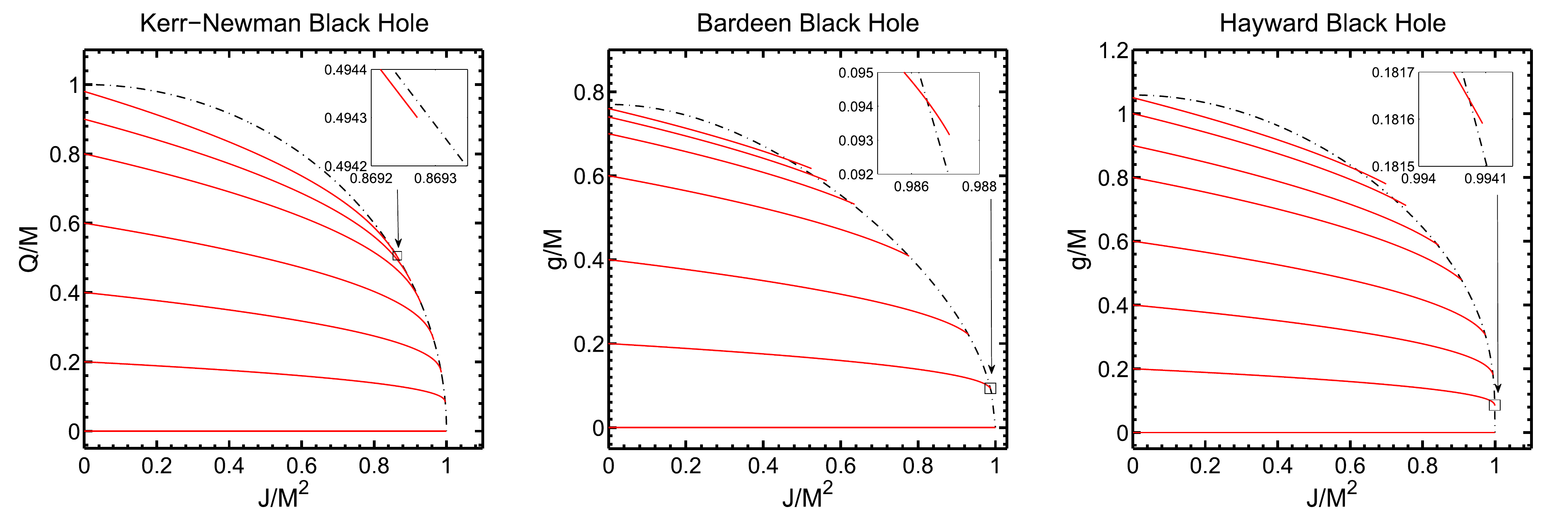}
\end{center}
\vspace{-0.5cm}
\caption{Tracks of the evolution of Kerr-Newman black holes (left panel), Bardeen
black holes (central panel), and Hayward black holes (right panel) as a 
consequence of the accretion process from a thin disk of uncharged gas. The 
black dashed-dotted line is the boundary separating black holes and horizonless 
states. The red curves cannot cross this boundary in the case of Kerr-Newman 
black holes (the accretion process cannot destroy the black hole), while they 
can do it in the case of regular black holes (the accretion process can destroy 
the black hole). See the text for details.}
\label{fig2}
\end{figure*}

It is even more interesting to note that it is probably not necessary an
ideal experiment to destroy a regular black hole, but a common astrophysical
phenomenon like the accretion process from a disk can presumably do the job.
In this framework, the gas falls onto the black hole as it loses energy and angular 
momentum. In the case of a thin disk on the equatorial plane, the gas reaches 
the innermost stable circular orbit (ISCO) and it then plunges quickly onto the 
compact object, which changes its mass $M$ and spin $J$ by~\cite{spin}
\be
M \rar M + \delta M \, , \quad J \rar J + \delta J \, ,
\ee
where
\be
\delta M = \epsilon_{\rm ISCO} \delta m \, , 
\quad \delta J = \lambda_{\rm ISCO} \delta m \, ,
\ee
$\epsilon_{\rm ISCO}$ and $\lambda_{\rm ISCO}$ are, respectively, the specific 
energy and the specific angular momentum of a test-particle at the ISCO, while
$\delta m$ is the gas rest-mass. Here, we still assume that the particles are 
neutral, so $Q \rar Q$ or $g \rar g$. Fig.~\ref{fig2} shows the evolution of our
black holes as a consequence of the accretion process from a thin disk (red solid
curves) for an initially non-rotating object and different values of the initial charge.
In these numerical calculations, we have used $\delta m = 10^{-6}$ or $10^{-7}$. 
The black dashed-dotted curves are the boundaries separating black holes from
horizonless states (naked singularity in the Kerr-Newman case, some kind of
regular solitons for the Bardeen and Hayward ones). As can be seen in the 
enlargement in the top right corner of every panel, in the case of the Kerr-Newman 
black hole, the red solid line cannot cross this boundary, even for high values of the 
electric charge $Q$, while that is possible for the regular black holes, even for relatively
low values of $g$ (and it becomes more and more easy as $g$ increases).

{\it Conclusions ---} 
Recently, there have been a lot of interest in the possibility of destroying the
event horizon of a black hole to violate the weak cosmic censorship conjecture. 
In the test-particle approximation, it seems like there are some orbits for which 
a small particle can plunge onto the black hole and destroy its event horizon. 
However, the allowed range of the parameters of the small particle is very 
narrow and of the same order of the expected corrections from the particle's 
self-energy, suggesting that the latter prevents the destruction of the black 
hole if properly taken into account. In this letter, we have considered the
possibility of destroying a regular black hole. As this object has no central 
singularity, it is not protected by the weak cosmic censorship conjecture.
We have presented two different examples (gedanken experiment and
accretion process from a thin disk) strongly suggesting that regular black holes
can be destroyed. Unlike the Kerr-Newman case, the parameter region
for which the test particle can destroy the black hole's event horizon is not
narrow, while the expected correction from the self-energy is of the same order
of the one in the Kerr-Newman spacetime. Moreover, it seems so easy to destroy such 
regular black holes that we do not need ideal experiments with particles 
of very specific $E$ and $L$, but the natural accretion process from a thin 
disk can do the job. Lastly, let us notice that our regular black holes violate 
the black hole's area theorem. For standard black holes, this theorem has 
important implications like the existence of an upper bound for the energy 
released in the collision of two black holes~\cite{haw}. In the case of regular 
black holes, such a bound does not exist and that may be tested by future 
detections of gravitational waves from the coalescence of two black holes.

%%%%%%%%%%%%%%%%%%%%%%%%%%%%%%%

\begin{acknowledgments}
%We would like to thank . . .  for . . .
This work was supported by the Thousand Young Talents Program 
and Fudan University.
\end{acknowledgments}

%%%%%%%%%%%%%%%%%%%%%%%%%%%%%%

\end{document}